\documentstyle[11pt,aaspp4,tighten]{article}
\def\rosat{{\sl ROSAT~}}
\def\asca{{\sl ASCA~}}

\newcommand\hii{H{\small II}}
\newcommand\hi{H{\small I}}
\newcommand{\as}{$^{\prime\prime}~$}

\slugcomment{}
\begin{document}

\title{\bf An Ultradeep High Resolution X-ray Image of M101:}
\title{\bf The X-ray Source Population in a Late-type Spiral}

\author{Q. Daniel Wang}
\affil{Dearborn Observatory, Northwestern University}
\affil{ 2131 Sheridan Road, Evanston,~IL 60208-2900}
\affil{Electronic mail: wqd@nwu.edu}

\author{Stefan Immler \& Wolfgang Pietsch}
\affil{Max-Planck-Institut f\"ur extraterrestrische Physik}
\affil{Postfach 1603, D-85740 Garching, Germany}
\affil{Electronic mail: simmler@mpe.mpg.de \& wnp@mpe.mpg.de}
\begin{abstract}

	We have studied the X-ray source population of the face-on spiral 
galaxy M101 (NGC 5457). Within a field of radius $17^\prime$ 
(36~kpc at the distance of 7.2~Mpc), covered by an ultradeep (229 ks) 
\rosat HRI 
image, 51 X-ray sources are detected with signal-to-noise ratios greater than 
3.5. About half of these sources are associated with the galaxy. The 
luminosity of these galactic sources individually ranges from $\sim 4\times 
10^{37}$ to $2\times 10^{39} {\rm~ergs~s^{-1}}$ in the 0.5-2~keV band. 
The average luminosity distribution of the sources can be characterized 
by a power law function: $d N/d L_{\rm x} = 9.5 L_{\rm x}^{-1.9}$ 
sources per $10^{38} {\rm~ergs~s^{-1}}$. 

Combined with archival data from the \rosat\ PSPC, the {\it Einstein} IPC, 
and the \asca\ GIS, we have examined spatial, spectral, and timing 
properties of the X-ray sources. In particular, we have explored the nature 
of various superluminous X-ray sources with luminosities significantly
greater than the Eddington limit ($\sim 2 \times 10^{38} {\rm~ergs~s^{-1}}$) 
for a $\sim 1.6 M_\odot$ object (neutron star). These X-ray sources, 
detected in various \rosat\ HRI and PSPC observations, are not transients
and appear to result from recent massive star formation in outer spiral arms. 
3 superluminous PSPC sources are associated with giant \hii\ complexes 
and are clearly resolved. 2 other superluminous RHRI sources 
are likely associated with shell-like supernova  (or more likely hypernova) 
remnants, which are known to be abnormally luminous in optical and/or radio. 
We further identify 2 superluminous sources, which all show highly-absorbed
 X-ray spectra and time variability during and/or between the observations, as
candidates for X-ray binary systems that contain black holes. 

	A comparison of 7 nearby spirals shows 
that their X-ray source luminosity distributions, 
normalized by total \ion{H}{1} masses, are very similar. But both the number of
superluminous X-ray sources and the total X-ray luminosity appear
to be correlated with the star forming rate  of a galaxy.

\end{abstract}


\section{ Introduction}

	Soft X-ray investigation of nearby galaxies is essential for our 
understanding of high energy phenomena and their relations to galaxy 
evolution. Such a study avoids much of the usual difficulty in determining
the distances to X-ray sources in our own Galaxy. For face-on galaxies 
at high Galactic latitudes, line-of-sight soft X-ray absorption is minimal, 
which can be particularly important for observing sources with soft X-ray 
spectra, such as supernova remnants (SNRs) and superbubbles. 
These galaxies also allow for rigorous cross-examinations of various 
galactic components and their interplay. Of course, the understanding of 
X-ray source populations in nearby galaxies is critical for correctly 
interpreting X-ray observations of distant galaxies.

	As a nearly face-on, grand-design spiral in a direction of 
exceptionally low foreground Galactic extinction (Table 1; Fig. 1), M101 is 
a prime target for characterizing the overall properties of
galactic X-ray sources. X-ray emission from this galaxy was first 
observed with the Imaging Proportional Counter (IPC) on board the 
{\sl Einstein} X-ray Observatory (McCammon \& Sanders 1984; Trinchieri, 
Fabbiano, \& Romaine 1990, TFR hereafter). The IPC observations show an 
apparent over-density of bright X-ray sources in the field of M101. 
The luminosity of these sources, if located at the distance of the galaxy,
is individually a few times $10^{38}-10^{39} {\rm~ergs~s^{-1}}$ in the IPC
0.16-3.5~keV band, significantly greater than the Eddington luminosity 
$\sim 2 \times 10^{38} {\rm~ergs~s^{-1}}$ for a $\sim 1.6 M_\odot$ object 
(e.g., neutron star). The presence of such superluminous 
sources (SLS) is also evident in the fields of several other nearby 
galaxies (e.g., Long \& Van Speybroeck 1983). Naturally, these 
sources have been suggested as X-ray binary systems containing stellar mass 
black hole candidates (BHCs). With the limited sensitivity and spatial 
resolution of the IPC observations, however, one cannot
rule out the possibility that some of the SLS are actually composites of 
multiple X-ray-emitting objects
(e.g., Fabbiano 1998). 

	Relatively recent observations from \rosat\ have provided new
opportunities to study X-ray sources in M101. Based on 
a 34.5 ks exposure with the PSPC (Position Sensitive Proportional Counter; 
Pfeffermann et al. 1987), Snowden \& Pietsch (1995) have shown that
a substantial fraction of the observed X-ray emission, primarily at 
$\sim 1/4$~keV, is unresolved, indicating the presence of the hot 
interstellar medium (ISM) in the galaxy. Using the same PSPC observation 
and a short 18.4 ks \rosat HRI (High Resolution Imager --- RHRI; David et al. 
1996), Williams \& Chu (1995) have further suggested that giant \hii\ 
complexes (GHCs) are responsible for several of the discrete X-ray sources 
detected in the field of the galaxy. These GHCs are recent starburst regions 
brighter in both optical and radio than the LMC 30 Dor complex, which contains 
various luminous X-ray sources: the stellar mass BHC LMC X-1, 
30 Dor nebula, LMC-2 superbubble, and several young SNRs (Wang \& Helfand 1991). 
These studies have demonstrated that M101 is an ideal site to 
examine the galactic-wide X-ray source population, not only under normal 
circumstances but also under various starburst conditions. 

	We have obtained new deep exposures of M101 with the \rosat HRI. 
Together with existing observations, we constructed an RHRI image with a total 
exposure time of 229 ks. This single ultradeep
image covers essentially the entire galaxy over a field of radius $R \sim 
17^\prime$ (36~kpc at the distance of M101; Table 1), 
which makes a population study of galactic X-ray sources particularly 
convenient. The good counting statistics of the image enables us to 
utilize the high spatial resolution capability ($\sim 5^{\prime\prime}$ 
FWHM on-axis) of the instrument, important for resolving X-ray 
emission both in the galaxy's central region and in GHCs which are typically 
located in outer spiral arms of the galaxy. The detection of point-like sources 
in the image reaches a limiting 0.5-2~keV luminosity of $\sim 4 \times 10^{37} 
{\rm~ergs~s^{-1}}$. The RHRI data also allow for time variability analysis on 
various time scales. To enhance the long-term
timing capability and to characterize spectral properties of the X-ray sources,
we have further analyzed archival data from both the \rosat\ PSPC and the \asca\ GIS.
(The SIS data from the same \asca\ observation, however, were severely corrupted 
because of bad telemetry and are not useful.) These X-ray data provide a
complementary spatial and spectral coverage over the entire galaxy.

	In this paper, we concentrate on discrete X-ray sources in the field 
of M101. We describe the X-ray observations and data 
reduction in \S 2, and catalog X-ray sources in \S 3. We 
examine X-ray properties of individual sources (\S 4) and, in particular, 
explore the nature of SLS (\S 5). We statistically 
characterize the flux distribution of the sources (\S 6) and its relation to
the overall X-ray emission from the galaxy (\S 7). We further compare our 
results with observations of X-ray source populations in
other nearby galaxies (\S 8). Finally in \S 9, we summarize our results 
and conclusions. In a subsequent paper, we will study the 
filamentary structure of the diffuse X-ray emission revealed in the RHRI image,
will make detailed comparisons with observations in other wavelength bands, 
and will discuss implications for the structure and evolution of the hot ISM 
in the galaxy.

\section {Observations and Data Calibrations}

	Table 2 presents a log of main X-ray observations used in the present
study. The nominal pointing directions of the \rosat\ observations were
identical: R.A., Dec. (J2000) = $14^h 3^m 9\fs6$, $+54^\circ 20^\prime 
24^{\prime\prime}$, about 0\farcm65 southwest of the nucleus of M101 
(Table 1). We constructed a single 229 ks RHRI image from 71 
individual observing segments included in the 4 observations. 
We aligned individual segments by matching the centroids of 12 brightest
X-ray sources in the field. No offset 
was found to exceed 6\farcs5. A comparison between the co-added radial 
intensity profiles of point-like sources within $\sim 12^\prime$ off-axis
before and after the bore-sight correction shows that the average point 
response function (PSF) is improved 
from 9\farcs5 to 7\farcs0 (FWHM). X-ray sources in the field are, however, 
not bright enough to allow for correcting potential errors 
related to the phase of the telescope wobble. 

	To obtain an accurate attitude solution of the RHRI image, we 
compared centroid positions of point-like X-ray sources 
(H1, H11, H13, H20, H38, H42, and H48; Table 3) with apparent optical 
counterparts listed in Guide Star Catalog (GSC\footnote{version 1.2, see 
document at 
${\rm http://www-gsss.stsci.edu/gsc/gsc12/gsc12\_form.html}$}) or in the 
APM catalog (Irwin, Maddox, \& McMahon 1994). We found a satisfactory position 
alignment by shifting the X-ray images 1\farcs7 to the west and 0\farcs7 to 
the north and by using a plate scale of 0\farcs499 per pixel (see also 
Wang 1995). 

	To reduce the non-cosmic contamination in the RHRI image, we selected 
only counts within the PHA channel range 2-10. Using a software provided by 
Steve Snowden, we further obtained both the effective exposure and
non-cosmic background maps of the co-added RHRI image. The 
background-subtracted and exposure-corrected intensity image is shown in 
Fig. 2.

Complementing the RHRI image, the PSPC observation provided a reasonable 
spectral resolution of $\delta E/E \sim 0.43 (0.93{\rm~keV}/E)^{0.5}$.
While both instruments were sensitive to photons in the 0.1-2~keV range,
the sensitivity of the PSPC was greater than that of the RHRI, 
particularly in the 0.1-0.3~keV band. However, the PSPC spatial resolution 
of about 0\farcm5 on-axis at $\sim 1$~keV was substantially lower than 
that of the RHRI. In the present study, we used only the inner portion of the 
PSPC observation (Fig. 3) 
to match the RHRI field of view. This avoided the complication that would be
caused by the instrument's supporting ribs at $\sim 20^\prime$ off axis.
Based on a comparison with optical counterparts, as in the case of the RHRI 
image, we obtained a satisfactory attitude solution for the PSPC observation:
5\farcs4 to the east and 9\farcs9 to the north. 

	The \asca\ GIS observation (Fig. 4) covered a broad energy range 
between 1-10~keV. The spectral resolution of $\delta E/E = 0.078 
(5.9 {\rm~keV}/E)^{0.5}$
was considerably better than the PSPC, although the spatial resolution was poor
with an on-axis FWHM of $\sim 1^\prime$ and a 50\% half power diameter of 
$\sim 3^\prime$. A shift of 1\farcm23 to the north led to a reasonably good
match of X-ray source centroids in the \asca\ and \rosat\ observations. 

\section{X-ray Source Detection and Catalogs}

	We searched for X-ray sources in both the PSPC and RHRI images.
Following a standard procedure, we conducted local (sliding-box) and map 
detections as well as final maximum likelihood (ML) analyses of individual 
sources. We chose the on-source detection aperture as the 90\% power-encircled
radius of the RHRI PSF, which is a function of off-axis angle. For the RHRI, 
the radius ranges from 9\as\ on-axis to 44\as\ at 17$^\prime$ 
off-axis. Both local and map detections were based on the RHRI image with 
bin size equal to 2\farcs5. To form a smoothed total background map,
we first re-binned the RHRI image by a factor of 8 after sources detected in the 
local detection had been removed. We then applied a median filter to the 
image with an effective smoothing area of 140\as$\times140$\as\
around each bin. This smoothing area was large enough for us to neglect 
statistical uncertainties in the background estimate ($\lesssim 2\%$). 
The background in source-removed regions was simply the median interpolation 
from their surroundings. This median filtering effectively removed small-scale 
($\lesssim 1^\prime$) features, while keeping the intensity gradient 
on larger scales. The resultant background map was insensitive to the exact 
flux threshold for the source removal and was later used to assess the flux 
limit of the source detection across the field (\S 6). 

	In the RHRI image, we found 51 X-ray sources with signal-to-noise 
ratios greater than 3.5 (Fig. 2). Properties of these 
sources are summarized in Table 3:
RHRI source number (col. 1); right ascension and declination 
of source centroid (cols. 2 and 3); 1$\sigma$ error radius (col. 4), including a 
systematic error of 3$^{\prime\prime}$ estimated from the position dispersion
in the attitude solution; signal-to-noise ratio of the 
detection (col 5); count rate (col 6) corrected for both vignetting and 
scattering for a point-like source; source energy 
flux (col 7) converted from the count rate (see below) and corrected for 
the Eddington bias (Hogg \& Turner 1998; \S 6); and flag on source variability 
and extent (col. 8). 

	The conversion from a count rate to an energy flux depends on the 
spectral shape of a source (Fig. 5). Foreground Galactic stars 
often have X-ray spectra which can be characterized as 
optically-thin thermal plasma with temperatures of a few $\times 
10^6$~K. AGNs in the \rosat\ band typically have a power law with 
a photon index of $\sim 2$ (e.g., Hasinger et al. 1998). 
For such spectra, a reasonably good approximation of the conversion is 
$4 \times 10^{-11} $$({\rm~ergs~cm^{-2}~s^{-1})/(counts~s^{-1}})$ 
(Fig. 5), which is adopted throughout the paper. The uncertainty should 
be less than a factor of $\sim 2$, except for sources located within or beyond 
dense clouds with foreground column densities $\gtrsim 2 \times 10^{21} 
{\rm~cm^{-2}}$. 

	Table 4 lists apparent optical counterparts of 31 X-ray sources. We
identified these optical counterparts from spatial cross-correlations of
the RHRI sources (Table 3) with objects included in Simbad, GSC, and APM as
well as the SNR catalog of M101 (Matonick, \& Fesen 1997; MF hereafter).
We adopted a matching radius as $2 \times \Delta_{err}$ (Table 3) to keep the
candidate of each source unique. Our ongoing spectroscopic observations 
of X-ray sources in fields of nearby galaxies show that the 
$B-R$ color of optical counterparts is discrimative: Bright ($B \lesssim 20$) 
AGNs essentially all have $B-R < 1$ (blue objects), while both galaxies and 
foreground normal stars typically have $B-R > 1$ (red objects). We
classify bright ($R \lesssim 17$ mag) stellar (point-like) 
red objects as Galactic stars, extended red 
objects as galaxies, and bright stellar objects with $B-R < 0.8$ 
as AGNs. Although blue objects could also be white dwarfs in our 
Galaxy, they would show very soft X-ray spectral
characteristics and would then be identified easily. 

	Table 5 lists 33 PSPC sources detected in 
the hard band (0.5-2~keV; PI channel 52-201; Snowden et al. 1994). A search 
in the soft band (PI channel 20-41, centered around 1/4~keV) did not 
yield any new source, except for a few weak peaks of apparently diffuse 
X-ray emission; the channels below 20 were not used to avoid the so-called 
ghost imaging problem. The hardness ratios are defined as HR1 $=
(hard-soft)/(hard+soft)$ and HR2 $= (hard2-hard1)/(hard2+hard1)$, where
$soft$ and $hard$ stand for the net source count rates in the soft and 
hard bands, while $hard1$ and $hard2$ are the split of 
a hard-band rate into two: channel 52-90 ($\sim 0.75$~keV) and 91-201 
($\sim 1.5$~keV). The unphysical values of HR1 $> 1$ for several sources
are a result of statistical uncertainties in the data, most serious in the 
soft band. HR1 is particularly sensitive to X-ray absorption, whereas HR2 to
the intrinsic X-ray spectral shape of a source. 

	Table 5 also includes the likely RHRI counterparts of individual PSPC
sources, based on their position coincidences within the detection apertures.
For ease of comparison, both PSPC and RHRI positions of individual sources
are marked in Fig. 3.  The RHRI image
is more sensitive to point-like sources than the much shallow PSPC obseration.
A considerable number of RHRI sources thus are not detected in the PSPC 
observation. Also with its limited spatial resolution, the PSPC was unable to 
resolve several X-ray sources of multiple components in the inner part  
($\lesssim 10^\prime$) of the galaxy. Conversely, 3 PSPC sources
failed to enter the RHRI source catalog (Table 3). P12, most likely originating
in hot ISM of NGC5455, has a very soft spectrum and
is apparently extended. The other 2 sources (P3 and P23) are 
all at large off-axis angles where the PSPC sensitivity is comparatively good.

\section{Properties of Individual X-ray Sources}

	In Tables 3-5, we have included various flags and comments on 
individual sources. In this session, we systematically describe our
spatial, spectral, and timing analyses of the sources. 

\subsection{Spatial Extent}

	To identify extended X-ray sources, we compared the count rates
detected within the 50\% and 90\% power-encircled radii
 around each sources in Table 3 and corrected for scattered fractions 
(i.e., 50\% and 
10\%). We flagged a source as being likely extended, if 
the difference of the two rates is greater than 3$\sigma$. In the calculation
of the count rate uncertainties, we quadratically added to the statistical 
error of each source a systematic uncertainty of 10\% the total
count rate (Table 3). This uncertainty was estimated from the count
rates of relatively isolated point-like sources (e.g., foreground 
stars; Table 4).  We confirmed the extendness by comparing the average radial 
intensity profiles of the sources with the expected PSFs.
As an example, we show the profile of H23 (M101 nucleus) in Fig. 6. The 
fit of the PSF to the profile is not acceptable ($\chi^2/n.d.f = 84.1/28$).
This extended morphology, together with the soft X-ray spectral characteristics 
of the source (Table 5), suggests that much of the X-ray emission 
from the nuclear region arises in diffuse hot gas which emits soft X-rays.

\subsection{Spectral Characteristics}

	The hardness ratios in Table 5 provide a useful spectral 
characterization of individual sources. Fig. 7 illustrates the 
dependence of HR1 and HR2 on X-ray-absorbing gas column density and 
on the parameters of the two spectral models, the power law function and 
the Raymond \& Smith optically thin 
thermal plasma. The former is typical for AGNs, while the latter is for such
galactic sources as stars, SNRs, and superbubbles. The absorption along a
line of sight to a foreground star should be less than the total Galactic 
gas column density (Table 1), while the absorption to an AGN should be a 
combination of the column densities in the Galaxy and in the M101, 
typically a few times $10^{20} {\rm~cm^{-2}}$ total. 
Complications could arise, though, if a PSPC source 
consists of mutiple components (RHRI sources). Within the PSPC source detection
aperture, the calculation of HR1, in particular, 
may be affected not only by the line-of-sight absorption but
also by the presence of diffuse hot gas, which typically has a very soft 
X-ray spectrum. For example, the abnormal hardness ratio of P21 
(both HR1 and HR2 have moderate values) may be a result of a spectral 
combination between a local diffuse soft X-ray emission enhancement
and an SNR embedded in a molecular cloud.
Such spectral consideration was included in
our classification of the X-ray sources (Tables 4 and 5). 

	The \asca\ GIS image (Fig. 4) maps out the relatively hard X-ray 
emission above
$\sim 1.5$~keV. Most noticeable is the absence of emission peaks 
at locations of several bright \rosat\ sources: P7 (star),  
P16 (nuclear region), P25 (GHC), P31 (star), and P32 (SNR), confirming that
these sources indeed have very soft X-ray spectra as expected. 
On the other hand,
the GIS image also shows regions of enhanced 
emission which do not correspond to any features seen in the
\rosat\ images. Particularly notable is the region northwest of the M101 nucleus.
 An enhancement of X-ray emission in the same region
is also present in the IPC image of the galaxy (TFR).
The nature of this relatively hard X-ray emission is unclear,
although it may represent a couple of X-ray binaries which were 
absent during the \rosat\ observations.

\subsection {Variability}

	We tested the variability of each RHRI source in Table 3, using the 
Kolmogorov-Smirnov statistic. To account for the variability in the 
non-cosmic X-ray background during the RHRI observations, we 
constructed a cumulative background count
distribution from counts in the RHRI image, excluding regions
within the 90\% power-encircled radii around individual sources. This 
background count distribution was normalized to the local 
background level around each source. The distribution plus a constribution 
from an assumed constant source flux was then compared with
the observed cumulative count distribution within the 90\% radii around 
the source. This comparison shows that 7 sources (H13, H19, H20, H28, H31, 
H32, and H45) are statistically inconsistent with being constant at the 
$3\sigma$ confidence (probability $P \gtrsim 0.9973$). 

	We also conducted $\chi^2$ tests for variability in the 
lightcurves of the RHRI sources. To achieve reasonable counting statistics, 
we blocked the lightcurves as 12~ks exposure intervals for sources
with individual count rates greater than $10^{-3} {\rm~counts~s^{-1}}$,
and as 20~ks intervals for fainter sources. We determined the time-dependent
background by normalizing the total background 
map of the RHRI image (\S 2) according to the total source-removed 
RHRI count rate in each exposure interval. 8 sources 
(H11, H13, H19, H20, H28, H31, H32, and H45) 
show significant varibilities, and their light curves are plotted in Fig. 8.

	Additional variability information is available from the 
comparison between the PSPC (hard band) to RHRI count rates 
in Tables 3 and 5. We assumed a conversion between 
the two count rates as $\sim 3$. But it could reach up to $\sim 3.5$ for very 
soft sources, such as foreground stars with little absorption, or down to 
$\sim 2.7$ for sources with strong absorption. With this uncertainty in the 
consideration, 4 PSPC sources (P8, P15, P18, and P28) and 1 
RHRI source (H12) appear to vary significantly 
between the PSPC and RHRI observations. This variability test, however,
is problematic for P8, since the point-like RHRI counterpart H10 is embedded 
within the diffuse X-ray-emitting GHC NGC5447. The count rate 
discrepancy between the PSPC and RHRI sources may be explained by the 
difference in the PSFs of the two instruments. The variabilities of other
sources are relatively convincing. In fact, during the RHRI observations,
both P18 and P28 also exhibited strong variability (Table 3), and P15 (H20) 
was marginally ($P=0.9951$) variable.

	A comparison with the results from the early IPC observations 
(TFR) is also interesting. 7 IPC sources (\# 1, 2, 4, 5, 6, 7, and 8) are
within the field of our consideration. IPC sources 4 and 6 disappeared during 
the PSPC and RHRI observations. The former was variable between the two IPC 
observations, and the latter coincided in position with a foreground star.

\section{Superluminous X-ray Sources in M101}

	12 RHRI sources (and more PSPC sources) have fluxes greater than 
$\sim 3 \times 10^{-14} {\rm~ergs~cm^{-2}~s^{-1}}$ in the 0.5-2~keV band 
(Table 3). If associated with M101, such a source has a luminosity of 
$\gtrsim 2 \times 10^{38} {\rm~ergs~s^{-1}}$ in the band; its bolometric 
luminosity could be much higher and exceeds the Eddington limit for a neutron 
star with a typical mass of $\sim 1.6 M_\odot$. Such sources are
thus superluminous (SLS or ``super Eddington'') and may be good BHCs. Of course, 
a fraction of these sources (e.g., H1, H6, H11, H20, and H48; Table 3), 
mostly projected in outskirts of
the galaxy, are just interlopers (stars, AGNs, and background galaxies; Table 4).
But, as we will show in the next session, there are definitely more
such X-ray sources in the M101 field than what is expected in a random field.

	SLS in M101 
apparently have several origins. 4 PSPC sources (Table 4), 
which may be classified as SLS, 
are clearly resolved by the RHRI: P8 into a point-like source (which is still 
superluminous though) and a diffuse component (NGC5447), P19 into 2 separate 
sources of comparable fluxes (2 SNR candidates; Wang 1999), P22 into 
apparently diffuse emission (NGC5461), and P25 into
multiple components (NGC5462). Thus at least a fraction of such sources, 
especially as seen at low spatial resolution, 
arises in shock-heated hot gas associated with SNRs and/or GHCs. 

	Two point-like SLS (H19, and H45) are 
good candidates for BHCs.  Both varied strongly during the 
RHRI observations (Table 3; Fig. 8). 
As expected, none of the two sources has an optical counterpart to the 
APM magnitude limits (see Note to Table 4). However, within its position 
uncertainty, H19 coincides with an unresolved optical emission line 
feature,  which has been classified as an SNR (Table 4), although their 
physical relationship is not yet clear. The HR1 values ($\gtrsim 0.8$; 
Table 5)
of the sources indicate large X-ray absorption along the lines of sight. 
Indeed, the positions of the sources
are all projected well within two outer spiral arms of the galaxy (Fig. 1). 
Therefore, the sources probably represent two high-mass X-ray binaries. 
	
\rosat\ observations of other nearby galaxies also suggest that 
SLS have various origins. A large fraction of them 
are very young SNRs, in which supernova ejecta are strongly interacting with 
circumstellar materials. Such sources 
include SN1986J ($\gtrsim2 \times 10^{39} {\rm~ergs~s^{-1}}$) in NGC891 
(Bregman \& Pildis 1992), SN1979C ($\sim1 \times 10^{39} {\rm~ergs~s^{-1}}$)
in M100 (Immler, Pietsch \& Aschenbach 1998), SN1978K
($1 \times 10^{39} {\rm~ergs~s^{-1}}$) in NGC1313 (Ryder et al. 1993), 
and SN1980K ($\sim3 \times 10^{39} {\rm~ergs~s^{-1}}$) in NGC6946 
(Canizares, Kriss \& Feigelson 1982). Another similarly luminous source 
in NGC6946 apparently coincides with a young SNR ($\lesssim 3500$~yrs; 
Schlegel 1994). It is hence conceivable that a fraction
of the SLS in M101 is associated with 
very young SNRs. Other unidentified superluminous 
X-ray sources tend to be associated with GHCs. One example
is the source ($\sim1 \times 10^{39} {\rm~ergs~s^{-1}}$) in NGC4631 
(Wang et al. 1995), an interacting galaxy with enhanced star formation. 
The source
is located in a GHC at the border of an \hi\ shell (Vogler \& Pietsch 1996). 
Since the source is found to be variable, it is most likely a high-mass 
X-ray binary. 

	It is thus plausible that most, if not all, SLS are related to 
the formation of very massive stars. In particular,
the presence of GHCs is a strong indication for massive starburst activities.
GHCs not only power supergiant bubbles of hot gas, but also
likely produce various extreme objects, such as black holes and hypernovae 
(e.g., Paczy\'nski 1998) that can naturally result in super-energetic 
remnants.

\section {Luminosity Distribution of X-ray Sources in M101}

	To determine the luminosity distribution of X-ray sources in M101, we 
considered various observational selection effects and accounted for
interlopers (e.g., foreground stars and background AGNs) in our
X-ray source list (Table 3).

	We first approximately corrected for
the Eddington bias, caused by the combination of uncertainties in 
the X-ray flux measurements and the steep flux 
distribution of the sources. We estimated the ``true'' (ML) 
flux as $S=0.5[1+(1-4\gamma/(s/n)^2)^{1/2}]S_o$ from the observed 
flux $S_o$ of a source (Hogg \& Turner 1998). This correction 
is significant only for sources near the detection threshold $s/n \sim 3.5$. 
We thus chose the slope of the flux distribution as 
$\gamma = 1.94$ (see below). The correction was applied to each 
source and to the flux limit of our source 
detection at each RHRI pixel. The flux limit varied across the 
field as a function of the PSF size as well as the background and exposure 
in the RHRI image. Fig. 9 shows the average radial dependence of the flux 
limit, compared with individual source fluxes. 

	We then estimated the contribution of interlopers in our source list, 
based on the differential flux distribution obtained from the \rosat\ deep 
surveys (Hasinger et al. 1988). The distribution as a function
of the source flux $S$ (0.5-2~keV; in units of $10^{-14} 
{\rm~ergs~cm^{-2}~s^{-1}}$) has a form 
$n_b(S)= n_{b,0} S^{-\gamma_b}$, where $n_{b,0} = 238.1$ and
$\gamma_b = 2.72$ for $S > 2.66$, or 111.0 and 1.94 for $S < 2.66$, 
respectively. We integrated the function to obtain the expected 
contribution above our
source detection flux limit at each pixel of the RHRI image. 
Fig. 10 shows the contribution 
as a function of galactocentric radius $R$ of M101. The excess of the
source density above the interloper contribution decreases with  
increasing $R$. In the $R \lesssim 12^\prime$ region,
where variation in the flux limit is relatively small (Fig. 9), the number of 
estimated interlopers is 19.7, and $21.3\pm6.4$ sources are expected 
to be associated with M101. Most interesting is an excess of sources with $S 
\gtrsim 3$ within the $R = 5^\prime-12^\prime$ annulus: 
8 such sources compared to the expected 2.6 interlopers. This excess is 
significant at the confidence of 99.8\%. 
In comparison, no such sources are at smaller $R$ in the RHRI image.
Between $R = 12^\prime-17^\prime$ the observed number of sources with $S 
\gtrsim 3$ is  4, compared to the expected 2.7. Thus the excess is 
statistically insignificant. Indeed, 3 of the 4 sources are 
identified (Table 4): H1 and H20 as foreground stars, and H51 as a background 
galaxy. While the excess of sources is most significant in the 
$R \lesssim 5^\prime$ region, they all have $S \lesssim 3$.
In contrast, no significant difference is found between the observed and the 
expected number of sources with $S \lesssim 3$ in the $R \gtrsim 5^\prime$
region. 

	We next parameterized the flux distribution of the M101 sources as a 
power law function: $n(S) = n_{o} S^{-\gamma}$. Using this function and the
above broken power law for interlopers, we conducted an ML fit 
(e.g., Cash 1980) to the fluxes of the 41 RHRI sources with $R \leq 12^\prime$.
This fit gave $\gamma = 1.9(1.6-2.4)$ (90\% confidence interval). 
The effect of the Eddington correction on $\gamma$ is considerable 
($\sim 15\%$), but is still smaller than the statistical uncertainty. 
As a function of $\gamma$, the normalization is $n_{o} = 105(79-128)$. The
fit is satisfactory with the $V_e/V_a$ statistic 
(Avni \& Bahcall 1980) equal to  0.51, compared to the expected value of
$0.5\pm0.045 (1\sigma$). Fig. 11 compares the flux distribution of the 
sources with the best-fit model. 

	The power-law function can be readily converted into the average 
luminosity distribution of M101 sources: 
$d N/d L_{\rm x} \equiv N_o L_{\rm x}^{-\gamma}$, where $N_o = 9.5(8.0-9.7)$ 
sources per $10^{38} {\rm~ergs~s^{-1}}$ in the 0.5-2~keV band. Including
the Poisson uncertainty in the total number of the sources, we obtain 
$N_o = 9.5(4.0-14)$. 

	The luminosity distribution may, however, vary with $R$. As noted 
earlier, there is a clear excess of $S \gtrsim 3$ sources in 
the $R = 5^\prime-12^\prime$ annulus. The number of lower 
flux sources is comparatively small. An ML fit to sources in the annulus only
gives $\gamma = 1.7(1.4-2.2)$, which is considerably smaller than the above 
value for 
the entire $R \leq 12^\prime$ region. In contrast, the central region contains
16 sources in the $S \sim 0.4-3$ range, but no source with $S \gtrsim 3$. 
The expected number ratio $N(S\gtrsim 3)/N(0.4 \lesssim S \lesssim 3)$ 
is 0.16, assuming $\gamma = 1.9$, and greater if $\gamma \sim 1.7$ is used. 
The probability for detecting no source in the $S \gtrsim 3$ range is thus
$\lesssim 0.073$. 
Unfortunately, both the flux range and the number statistics of the sources 
in the central region are too small to constrain effectively an independent
luminosity distribution. Nevertheless, the above analysis indicates
that the central region intrinsically lacks SLS and that
the luminosity distribution depends on $R$. This radial dependence may be 
related to the large metallicity gradient observed in the galaxy (a factor
of 10 decrease from the galactic center to outskirts; e.g., 
Kennicutt \& Garnett 1996; Torres-Peimbert, Peimbert, \& Fierro,  1989).
Massive stars with low metallicity may be relatively inefficient in losing 
mass, and may more likely end their life as hypernovae 
(e.g., Paczy\'nski 1998) and form black holes than stars with high 
metallicity. 
This metallicity effect may partly explain why no persistent SLS have been 
observed in such early-type spirals as M31 and our 
own Galaxy, both of which have relatively high metal abundances (\S 8).
 
\section{Overall X-ray Emission from M101}

	In addition to the detected sources, as listed in Tables 3 or 5, 
there are large amounts of low-surface brightness X-ray emission associated 
with M101. Fig. 12 shows an RHRI intensity map that has been adaptively 
smoothed to emphasize this emission. After both excising individual sources 
within their 90\% power-encircled radii and subtracting a background 
calculated in the annulus of $R = 12^\prime - 15^\prime$, we estimated the 
excess of the ``diffuse'' X-ray emission as 0.028 
${\rm~counts~s^{-1}}$ in the $R \leq 5^\prime$ region and 0.02 
${\rm~counts~s^{-1}}$ in the $R = 5^\prime - 12^\prime$ annulus. In 
comparison, 
the total count rate of the sources in the same regions are 0.0052 and 0.012 
${\rm~counts~s^{-1}}$, including the contribution from interlopers.
Clearly the fractional contribution from the
sources is substantially greater in the annulus
than in the central region, although our source detection flux limit is 
higher in the former than in the latter. The total 0.5-2~keV 
luminosity of the diffuse emission within $R = 12^\prime$ is $\sim 9 
\times 10^{39} {\rm~ergs~s^{-1}}$.

	Can discrete sources just below our source detection limit explain 
the apparently diffuse X-ray emission? Indeed, a modest extrapolation of the 
average luminosity distribution of M101 sources (\S 6) may reasonably account for
the excess flux in the $R = 5^\prime - 12^\prime$ annulus. To explain the 
flux in the $R \leq 5^\prime$ region, however, $\gamma 
\gtrsim 2$ is required. For $\gamma = 2$, for example, the extrapolation needs to 
reach $L_{\rm x} \sim 7 \times 10^{34} {\rm~ergs~s^{-1}}$, which is substantially
smaller than the typical luminosity of an X-ray binary system 
($L_{\rm x} \gtrsim 10^{36} {\rm~ergs~s^{-1}}$). As argued by Snowden \& Pietsch
(1995), other less luminous types of X-ray sources (e.g., M dwarfs and 
white dwarf accretion binaries) are also unlikely to account for the
bulk of the X-ray emission from the central region of the galaxy.
Therefore, the central region probably contains truly diffuse X-ray emission.

	To further explore the nature of the X-ray emission from the central
region, we studied spectral data from both the PSPC and GIS observations.
We extracted the data from the $R \leq 5^\prime$ region and estimated the 
background contribution from an annulus of $R \sim 12^\prime - 15^\prime$.
Because the limited spatial resolution of the data, we did not attempt to 
remove the sources. We obtained satisfactory joint fits to the PSPC 
and GIS spectra (Fig. 13), assuming the two-component models as listed in 
Table 6. Both models give a 0.1-10~keV luminosity as $\sim 2.0 \times 
10^{40} {\rm~ergs~s^{-1}}$. The high temperature thermal plasma 
(or the power law component) alone predicts an RHRI count rate of $\sim 0.01
{\rm~counts~s^{-1}}$, or about twice the total detected source count rate in 
the region. Thus it is reasonable to assume that this component arises in 
discrete sources, most of which may be X-ray binaries in the luminosity 
range of $10^{36}- 10^{38} {\rm~ergs~s^{-1}}$ and tend to have relatively
hard X-ray spectra. The low-temperature component 
is likely dominated by truly diffuse hot gas. The characteristic
temperature is even lower than that (0.26-0.60~keV)
of diffuse hot gas observed in the Large Magellanic Cloud (Wang et al. 1991; 
Tr\"umper et al. 1991; Snowden \& Petre 1995).  This is, at least partly, 
due to the high gas column density ($N_{HI} \sim 6
\times 10^{20} {\rm~cm^{-2}}$ in our Galaxy alone) toward the LMC; very soft
($\lesssim 0.3$~keV) X-rays  are severely absorbed, which 
biases the detection to gas at temperatures $\gtrsim 0.3$~keV
in the Cloud. There might be an even cooler (a few times $10^5$~K) 
gas component in M101.
The best-fit $N_H$ in Table 6 is considerably smaller than the expected 
Galactic column density (Table 1). This is an indication for the presence
of a very soft component in the spectral data (Wang et al. 1995), although
the quality of the data used here is not good enough to 
quantify this spectral component.

We have further found that the diffuse X-ray emission is
strongly correlated with \hii\ regions in M101. Thus as in the LMC (Wang 
et al. 1991), we expect that the X-ray emission is dominated by hot gas 
heated by a combination of supernova blastwaves and stellar winds from 
massive stars. We will explore the implications of this correlation 
further in a later paper.

\section{Comparison with Other Nearby Spiral Galaxies}

	We concentrate here on the comparison of M101 with 6 other
nearby spiral galaxies that we are familiar with.
Table 7 lists the salient parameters of these galaxies, arranged according to
their far-infrared (FIR) luminosities ($L_{\rm FIR}$). The galaxies cover 
a good range of 
morphological type, \ion{H}{1} mass, and star forming 
rate. The X-ray results are all based on \rosat\ observations. 
Because we are mostly interested in the overall galactic X-ray properties, 
the contributions from nuclear X-ray sources are excluded from both 
Table 7 and Fig. 14, where the X-ray source luminosity distributions
of the galaxies are compared. In addition, SN1979C of M100 is not included. 
This bright SN ($L_{\rm x}=10^{39} {\rm~ergs~s^{-1}}$) 
would have been detected in all other galaxies in many wavelength regimes. 
Furthermore, all identified interlopers are removed for each galaxy, 
according to the referenced papers (Table 7). But the data quality
and analysis differ significantly from one galaxy to another, which  
might cause errors in our calculation of the luminosity distributions, 
particularly in the lowest or the highest luminosity bins. 
The lowest luminosity bins 
are affected chiefly by uncertainties in source detection thresholds
and in various source confusions, most seriously in PSPC observations,
while the highest luminosity bins by the uncertainty in removing 
interlopers. The intermediate luminosity bins should be relatively reliable.
The galaxies seem to have very similar \ion{H}{1} mass-normalized X-ray 
source luminosity distributions (within current measurement errors), although the 
high luminosity cutoffs can be very different.

	Both the X-ray luminosity and the number of SLS 
 appear to be correlated with its {\sl total} star forming rate, as 
traced by the $L_{\rm FIR}$ of a galaxy (Table 7). 
The three galaxies (M33, M31, and NGC253) with the lowest $L_{\rm FIR}$ also have
the lowest $L_{\rm x}$ and $N_{\rm SLS}$, whereas the three galaxies 
(M51, M100, M83) with the highest $L_{\rm FIR}$ also have the highest $L_{\rm x}$ 
and $N_{\rm SLS}$. In terms of these properties, M101 is just intermediate 
among the galaxies listed in Table 7. 

\section {Summary}

	We have carried out a comprehensive study of X-ray sources
in the field of M101. We detected 51 sources in an ultradeep RHRI image 
and additional 3 sources in a PSPC observation. We have examined spatial, 
spectral, and timing properties of these sources as well as characteristics 
of their optical counterparts (magnitude, color, and extent).
We have also investigated the statistical properties of the sources and 
their relationship to the overall X-ray emission from the galaxy.
Furthermore, we have compared this study with works on other 
nearby galaxies. The major results and conclusions are as follows:

\begin{itemize}
\item  Significant X-ray emission is detected from M101 over a region 
of $R \sim 25$~kpc from the galaxy's nucleus. The total luminosity is $\sim 1
\times 10^{40} {\rm~ergs~s^{-1}}$ in the 0.5-2~keV band. 

\item  16 sources are identified as interlopers: 8 as
foreground stars, 3 as background galaxies, and 5 as AGNs (Tables 4 and 5). 
Unidentified interlopers ($\sim 10$) are most likely AGNs, which are 
relatively faint in optical. 

\item About 25 sources are associated with M101, which 
are in the luminosity range of $L_{\rm x} \sim (4-200) \times 
10^{37} {\rm~ergs~s^{-1}}$. 

\item  The 5 brightest GHCs in M101 all show strong X-ray 
emission (Table 5): 4 of them (NGC5447, NGC5455, NGC5461, and
NGC5462; Fig. 12) are clearly resolved, which most likely contain hot gas 
emitting diffuse soft X-rays; The emission from NGC5471 is likely dominated by a 
hypernova remnant (Wang 1999), which is also unusually bright in optical 
and radio (Yang et al. 1994; Chen \& Chu 1999). The diffuse emission
from the GHCs will be discussed further in a separate paper.

\item  Additional 4 position coincidences are found between X-ray sources and 
previously classified SNRs in the galaxy. Some of these coincidences, 
however, may result from superpositions by chance. The observed X-ray 
luminosities, if arising in blastwave-heated gas, suggest 
that these remnants originate in explosions much more energetic than typical 
supernovae. A thorough discussion of such remnants in M101 has been given by Wang 
(1999).

\item SLS, with luminosities significantly greater 
than the Eddington limit for a neutron star, are located exclusively in outer 
regions ($R \gtrsim 10$~kpc) of the galaxy. We identify 
2 variable SLS as candidates for X-ray binary systems 
with black holes. These sources show highly-absorbed 
X-ray spectral characteristics and are apparently embedded in spiral arms, 
indicating massive star origins.

\item  The average luminosity distribution of M101 sources is parameterized as a 
power law function: $d N/d L_{\rm x} = N_o L_{\rm x}^{-\gamma}$ 
with $\gamma = 1.9(1.6-2.4)$ and $N_o = 9.5(4.0-14)$ sources per $10^{38} 
{\rm~ergs~s^{-1}}$ (90\% confidence intervals). However, the 
luminosity distribution may be a function of galactocentric radius. 
In particular, X-ray sources in the central region appear to have an upper
luminosity cutoff at $L_{\rm x} \sim 2 \times 10^{38} {\rm~ergs~s^{-1}}$. 

\item  The detected sources account for only $\sim 16\%$ of the X-ray emission
observed in the central $R \lesssim 10$~kpc region. A substantial fraction 
of the residual emission likely arises in the hot ISM at a characteristic 
temperature of $\sim 2 \times 10^6$~K.

\item  Comparison of X-ray properties of 7 nearby spirals suggests that
both the X-ray luminosity and the number of SLS are
correlated with the galactic star forming rate. However, except for 
the different upper luminosity cutoffs, nearby spirals seem to 
have very similar X-ray source luminosity functions (after being 
normalized by the total \ion{H}{1} masses of the galaxies).

\end{itemize}

\acknowledgments
We thank the referee for comments that led to improvements of the presentation of 
the  paper. This research made use of various online services and databases 
(e.g., HEASARC website, Simbad, NED, APM, and DSS). Q.D.W. is supported by
NASA (grant NAG 5-3414 and NAG5-6413). The ROSAT project at MPE is supported 
by the German Bun\-des\-ministe\-rium f\"ur Bildung, Wis\-sen\-schaft, 
For\-schung und Tech\-no\-lo\-gie (BMBF/DLR) and the 
Max-Planck-Gesel\-lschaft (MPG).

\clearpage

\begin{figure} \caption{Optical image of M101 from Palomar Sky Survey. RHRI 
sources ($\Box$) and three additional sources ($\Diamond$), detected 
only in the PSPC observation, are marked. 
\label{fig1}}
\end{figure}

\begin{figure} \caption{RHRI intensity contour map of M101 and 
RHRI sources. The map is adaptively smoothed with a Gaussian with size
adjusted to achieve a constant signal-to-noise 
ratio of $\sim 4$. The contours are at 1.4, 2.1, 3.3, 6.5, 16, and 39 $\times 
10^{-3} {\rm~counts~s^{-1}~arcmin^{-2}}$; the lowest level is about 3$\sigma$ 
above the local X-ray background. Both positions ($+$) and 
catalog numbers of the sources (Table 3) are marked. 
\label{fig2}}
\end{figure}

\begin{figure} \caption{PSPC 0.5-2~keV band image of M101. The PSPC 
intensity map is smoothed in the same way as the RHRI map in Fig. 2, and the
contours are at 1.2, 1.8, 2.9, 5.6, 14, and  33 $\times 10^{-3} 
{\rm~counts~s^{-1}~arcmin^{-2}}$. Both positions ($\times$) and 
catalog numbers of PSPC sources (Table 5) are marked. For ease of 
comparison, the positions of RHRI sources ($+$) are also included. 
\label{fig3}}
\end{figure}
 
\begin{figure} \caption{1.5-10~keV band image of M101 from the \asca\ GIS. The 
GIS intensity map is smoothed in the same way as the RHRI map in Fig. 2, and the
contours are at 2.0, 2.6, 3.5, 4.7, 6.2, and 8.0 $\times 10^{-4} 
{\rm~counts~s^{-1}~arcmin^{-2}}$. The positions of RHRI sources ($+$) are marked. 
\label{fig4}}
\end{figure}

\begin{figure} \caption{Conversion factor (in units
of ${\rm 10^{-11} ergs~cm^{-2}~s^{-1}}/$${\rm~counts~s^{-1}}$) of 
an RHRI count rate to an unabsorbed (absorption-free) flux in the 0.5-2~keV band
as a function of 
model parameters. The thick curves are for power law with photon index 
of 1 (solid), 2 (dotted), and 3 (dashed), while the thin curves are for
Raymond \& Smith plasma with $kT = 0.2$ (solid), 0.4 (dotted), and 1~keV 
(dashed), respectively.
\label{fig5}}
\end{figure}

\begin{figure} \caption{Radial RHRI intensity profile around the M101 nucleus, 
compared to the PSF of the instrument (the solid curve). 
\label{fig6}}
\end{figure}

\begin{figure} \caption{Dependency of PSPC hardness ratios on
X-ray-absorbing gas column density and on spectral model parameters. 
The set of dashed curves is for HR1, while the set of solid curves is for 
HR2. In (A), the three curves (from the bottom to the top) of each set 
are for the power law model with photon index equal to 1, 2, and 3. In (B),  
the 5 solid curves (from the bottom to the top) are for the Raymond \& Smith
plasma with temperature equal to 0.1, 0.2, 0.5, 1, and 2~keV, while
the order for the 5 dashed curves are 0.1, 0.2, 2, 1, and 0.5~keV.
The metal abundances of both X-ray-emitting 
and -absorbing materials are assumed to be the solar.
\label{fig7}}
\end{figure}

\begin{figure} \caption{Light-curves of 9 RHRI sources which show significant
variability. The error bars of the data points are at the 1$\sigma$ confidence.
The reduced $\chi^2$ value as well as the variability significance and source
number are given on the right side of each panel. 
The dotted lines represents the mean count 
rates over the four individual observation blocks, which are separated by 
several months, while the dashed line indicates the mean count rate over the 
complete observation. 
\label{fig8}}
\end{figure}

\begin{figure} \caption{Comparison of RHRI source fluxes and the 
detection limit. The solid curve represents the
flux limit calculated with an azimuthally-averaged radial background 
intensity profile, while the dashed curves assume the background profile
plus or minus 3$\times$rms.
\label{fig9}
}
\end{figure}

\begin{figure} \caption{Number distribution of X-ray sources. The data 
points and 1$\sigma$ error bars are 
from the detected RHRI sources (Table 3). The solid curve shows the
estimated contribution from interlopers. The dashed curve represents
the contribution that assumes both an average flux distribution of sources 
within $R = 12^\prime$ and a uniform surface source density.
\label{fig10}
}
\end{figure}

\begin{figure} \caption{X-ray flux distribution of RHRI sources within 
$R = 12^\prime$. 
The units of the source density is the number of sources per 
$10^{-14}$ ${\rm~ergs~cm^{-2}~s^{-1}}$ (0.5-2~keV) and per degree$^2$. 
The horizontal bars represent the flux ranges of individual data point, 
whereas the vertical bars represent the square root of the observed source 
numbers (chosen to be 6, except for the highest flux range). The lower flux 
range extends to the minimum detection limit of the 
field. The upper limit of the highest flux range is arbitrarily selected 
to be the highest observed source flux; thus the density of the range 
is likely overestimated. The ML fit was performed on the data of individual 
sources without the binning. The best-fit model (M101 sources plus 
interlopers) is shown as the solid (intrinsic) and dashed (flux 
limit-corrected) curves. The thin curves represent the best-fit 
contribution from M101 only.
\label{fig11}
}
\end{figure}

\begin{figure} \caption{RHRI intensity map of the inner part of M101, 
adaptively smoothed with a constant signal-to-noise ratio of $\sim 6$. 
The contours are at 1.1, 1.3, 1.7, 2.2, 3, 4, 6, 12, 24, 48, and 100 $\times 
10^{-3} {\rm~counts~s^{-1}~arcmin^{-2}}$.
\label{fig12}
}
\end{figure}

\begin{figure} \caption{PSPC and GIS spectra of the inner 
$R \leq 5^{\prime}$ region around the nucleus of M101. The 
histograms represent the best-fit two-temperature thermal 
plasma model (Table 6), while the lower panel shows the residual of the fit.
\label{fig13}
}
\end{figure}

\begin{figure} \caption{A comparison of X-ray source luminosity 
distributions of nearby spirals (see also Table 7).
\label{fig14}
}
\end{figure}
\end{document}